# PROTOTYPE SMART HOME ENVIRONMENT WITH BIOFEEDBACK

1st Azmyin Md Kamal
Graduate Student,
Department of Mechanical Engineering
University of Louisiana at Lafayette
Lafayette, USA
c00441440@louisiana.edu

2nd Mushfiqul Azad
Graduate Student,
Department of Mechanical Engineering
Bangladesh University of Engineering and Technology
Dhaka, Bangladesh
azad.musfiqul@hotmail.com

3rd Sumayia Jerin Chowdhury
Department of Textile Engineering
Ahsanullah University of Science and Technology.
Dhaka, Bangladesh
tushin32@gmail.com

*Abstract*— In this paper we present a prototype of a smart home system which can actuate different peripherals based on the emotional "arousal" level of an user. The system is comprised of two embedded subsystems named "Wearable" and "Benchtop" which communicates with one another over UPD/IP protocol. The Wearable unit can differentiate the emotional arousal into three distinct classes (Normal, Medium and High) based on physiological data whilst the Benchtop unit can display different colors on a 16 digit NEOPIXEL ring and, play tones to emulate actuation of peripheral devices in the smart home environment. Experiments with three video clips were performed which showed that the system can classify emotional arousal with an average accuracy of 41%. An FSM model of the Benchtop unit was created using Ptolemy II which showed the model to be fully deterministic and robust to communication disruption between the two units. The proposed project will add a new paradigm in smart home and IoT research by incorporating emotional feedback to automatically adjust the indoor environment for greater comfort, ease of living and in-home assisted ambulatory care for the residents.

I. INTRODUCTION

*A. Background*

The dawn of the 21st century has seen a tremendous increase in the research, development and deployment of "smart environments" defined as "a system that is able to acquire and apply knowledge about an environment and to adapt to its inhabitants in order to improve their experience in that environment"[2]. The type of data acquired for the purpose of smart environment depends heavily upon the use case. For instance, the Georgia Tech's AwareHome project [3] uses IoT technology to conduct investigations on developing innovative sensing and control infrastructure for offering better health and wellness services to the residents with specific focus on improving the independent living and tele-health care for senior citizens [2,3]. In parallel to such developments, wearable sensors for measuring physiological parameters have been appearing extensively in "affective computing" research since 2000 [2-4]. In this regard, Schmidt el at in [1] gives a broad overview of measuring heart rate, respiration rate, oxygen content, galvanic skin response, skin temperature, movement using wearable sensors and/or smartphones for mapping people's emotion to Russel's circumplex diagram. In this regard, Zenonos el at in [5] used a wearable sensor from Toshiba to demonstrate a categorical affect recognition algorithm (i.e quantifying mood) to map a person's perceived emotion to eight different emotions and moods categories viz. "excited, happy, calm, tired, bored, sad, stressed and angry". Pham el at in [11] developed CoSHE, a cloud based Smart Health Environment system for delivering in-home health care for elderly citizens. This project consists of four major components: a smart home setup, a wearable unit, a private cloud infrastructure and a home service robot.

However, a common trend observed in the research highlighted above is that the wearable units are designed only to collect physiological data while a full spec computer processes the data and off-loads them to an external server for researchers and healthcare professionals to review. To the best of our knowledge, only the work in [11] had attempted to make a "smart health environment" which had an additional feature to control a mobile robot. This robot interacted with a person in response to certain health related events detected from user habits and their physiological signals. To the best of our knowledge we did not find evidence of any literature which attempted to classify emotions in the embedded system level. All the projects mentioned above utilized powerful computers equipped with advanced machine learning algorithms for feature recognition and emotion classification.

*B. Problem Statement*

Motivated by the lack of research to classify emotion in the embedded systems level and the lack of utilization of physiological signals in a smart home systems, the goal of this project was to build a prototype system which would be capable of determining the user's emotional arousal from a set of physiological signals and actuate some peripherals in real time without the aid of powerful computers and human-in-the-loop systems.

*C. Contribution*

The contributions of this research endeavor are as follow

a. Successfully demonstrated a *proof of concept* of a smart home environment which can autonomously actuate peripherals based on biofeedback from an user.
b. The system can classify arousal to three distinct classes with an average accuracy of 41%
c. Successfully demonstrated the deterministic nature of the Benchtop unit via a Ptolemy II simulation.

## II. EXPERIMENTAL SETUP

### A. Hardware and Software Setup

Figure 1. shows the hardware setup which comprises of two separate embedded systems dubbed "Wearable" (left) and "Benchtop" (right) respectively. The function of the "Wearable" is to collect physiological data and predict the emotional state of the user whilst the "Benchtop" unit functions as a proxy for the smart home. Each time a new class data is received, the Benchtop unit actuates its peripherals to reflect the class of arousal received. Note that, the USB cables shown in the Figure 1. are only for providing power to the embedded units, communication between the Wearable and Benchtop units occurs exclusively over Wi-Fi.

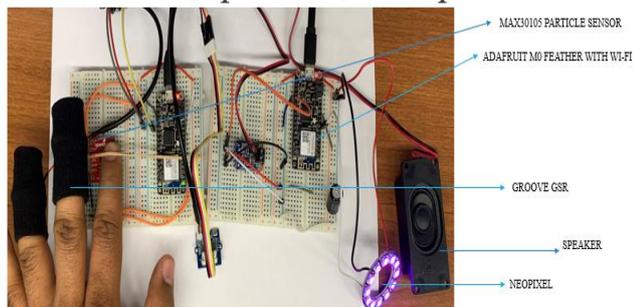

*Fig 1. Hardware setup*

The microcontroller (MCU) boards used in this project is the Adafruit's Feather M0, an ARM based prototyping platform composed of an ATSAMD21G18 ARM Cortex M0 processor, clocked at 48 MHz and running at 3.3V logic. It is capable of 12-bit ADCs and comes with native support for SPI, I2C and UART protocols. Wireless communication between the two units were established using the onboard Microchip Technology's ATWINC1500 Wi-Fi radio module which is an IEEE 802.11 b/g/n IoT Network Controller operating at 2.4Ghz channel. The Wi-Fi radios came soldered on the M0 boards and has an extensive software library which made it very easy get them connected to an Ad-Hoc network in the development phase. In the Wearable unit, a Sparkfun's MAX30105 particle sensor board and Seeed Studio's Groove GSR module are installed to evaluate heart rate and skin conductance, respectively. Adafruit's NEOPIXEL digital RGB LED strip was chosen to produce color actuation whilst a 4W low power speaker coupled with Adafruit's PAM8320A 2.5W Class D mono amp was chosen for playing different musical tones in the Benchtop unit. For programming the MCUs, open-source Arduino IDE was used and for capturing serial data, a SSH client PuTTY was used. Finally, The CPS model for the benchtop unit was simulated in Ptolemy II [1] to test its deterministic design. Details of this simulation is presented in Section V.

### B. Heart Rate and GSR

A brief introduction to Heart Rate and Galvanic Skin Response is presented in this section.

We begin by defining Heart Rate (HR) as the "speed of the heartbeat measured by the number of contractions (beats) of the heart per minute (bpm)" [8]. HR was measured using an optical technique known as Photoplethysmography (PPG). This non-invasive optical measurement technique involves illuminating tissue cells (e.g skin) with an IR beam and then measuring the small changes in the light intensity, associated with the changes to the blood volume, caused by the rhythmic cardiac cycle. "The PPG signal is primarily categorized into two components, a pulsatile ('AC') waveform attributed to cardiac synchronous changes in the blood volume with each heartbeat. The second category contains a slowly varying ('DC') baseline with various lower frequency components attributed to respiration, sympathetic nervous system activity, and thermoregulation".[7] MAX30105, the sensor used to implement PPG technology, is a reflectance type pulse oximeter (see Figure 2). During each sampling window, MAX30105 pulses its IR led and captures the reflected light which is proportional to the blood volume variation. During our experiments, we observed that the use of Infrared led gave the good result though it is possible to use the green led for more accurate reading [9]. Once the signal is recorded, the Penpheral Beat Amplitude (PBA) was used to attenuate the dominant DC component and boosts up the pulsatile AC component. After signal processing, the software automatically records each time the AC component cross a known threshold and then computes the Heart Rate in Beats Per Minute (BPM). The HR detection frequency was approximately 1Hz.

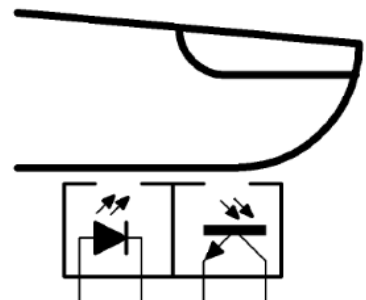

*Fig 2. Reflectance PPG [source: Internet]*

Galvanic Skin Response or Skin Conductance is the physiological response triggered by the sympathetic nervous system observed in a person in response to stressful or emotionally arousing stimuli. "Galvanic skin response directly correlates to the sympathetic nervous system activity and provides a powerful tool for monitoring arousal and certain aspects of the autonomic regulation." [12]. In stressful or arousing situation, the sweat glands are triggered and becomes more active. They secrete moisture (sweat) through the pores which directly correlates to the perceived stressful/arousing stimuli [7]. Sweat is a weak electrolyte and a good conductor, thus galvanic skin response is measured by applying a low voltage electric current to the skin and measuring the change in resistance which can then be translated to conductance. Figure 3 shows the palmar region of the hand where the test leads of a GSR sensor should be placed for best results.

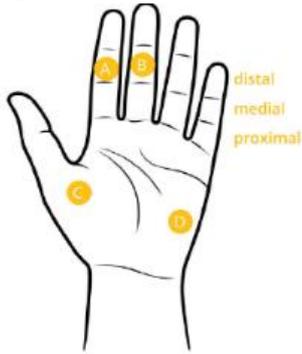

*Fig 3. Optimal GSR led locations [10]*

The Groove GSR module used in this experiment came with its own signal capture and filtration module. For increased accuracy we used a convolution filter with a window of 8 samples per successive heartbeat detections. This approach ensured that the GSR data was always synched with respect to each successive heartbeat detections. Synchronicity of physiological sensors in an important issue which can cause misclassification if not enforced in the hardware level [7,12].

*C. Assumptions*

The following assumptions were taken during our experiment
1. GSR response linearly ranged between 0 – 25 Microsemens and was unaffected due to environmental conditions.
2. Heart Rate data were considered relevant only between a range of 60 – 120 BPMs
3. Self-reported emotion arousal is sufficient as ground truth [12].

*D. Experiment Methodology*

The experiment was conducted in a lab environment under ambient temperature and lighting condition. At first the participant wore the GSR leads and placed his finger securely on the MAX30105 module. He was then prompted to commence data recording via screen prompt. Three video clips from the popular YouTube social media platform were chosen to generate emotion elicitation in the participant as video clips are commonly used for emotion elicitation in affective computing [4]. Each video clip was played for four minutes, two times. In first round, for every 15 seconds, the user self-reported his emotional arousal level in three classes [Normal, Mild and High]. In second pass, for each of the known 15 seconds intervals, the Wearable unit predicted perceived emotion under the same categories mentioned above. This was achieved by feeding the most recent BPM and GSR data into a linear ladder logic which assigned "weights" to the prediction. For example, if the heart rate (HR) was between 60-85 BPM but the GSR value was between 15-20 uS, the prediction was given more weight towards MILD than NORMAL. We justify this by stating that, it is possible for a person to have a higher arousal without having an increased heart rate [4,12]. During the automatic emotion prediction rounds, the Wearable issued an unique character byte over the Wi-Fi network towards the Benchtop unit. The Benchtop unit receiving this unique character, would then actuate the connected peripherals as shown in Table 1.

Table 1. Response to different character bytes

| *Arousal Level* | *Associated Character* | *Response from the Benchtop unit* |
|---|---|---|
| Normal | 'A' | Green color in NEOPIXEL, Tone 1 |
| Mild | 'B' | Orange color in NEOPIXEL, Tone 2 |
| High | 'C' | Red color in NEOPIXEL, Tone 3 |
| Invalid | 'D' | White color in NEOPIXEL, no Tone |
| BownOut | 'E' | Magenta color in NEOPIXEL, no Tone |

Apart from showing responses to the three-arousal level, our system also has two additional responses dubbed "BrownOut" and "Invalid". If the emotion data received did not match any character that maps the arousal class, the Benchtop unit enters "Invalid" state and sets the NEOPIXEL to white color. Additionally, if there is a drop of data stream between the Wearable and Benchtop units for more than 10 seconds, the Benchtop enters the "BrownOut" state, where it sets the NEOPIXEL to magenta color and waits indefinitely until another valid input is received.

### III. CPS MODEL

*A. Ptolemy II model*

In this project the Benchtop unit is a Discrete Event Finite State Machine whilst the Wearable is a Hybrid System. Figure 4. shows the high-level CPS model for the system.

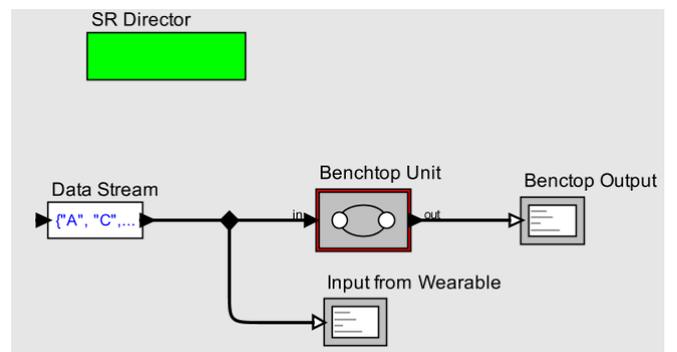

*Fig 4. CPS Model of the project*

The model is composed and simulated using Ptolemy II [1]. For avoiding complexity with Hybrid modelling, we have considered "Data Stream" sequence actor to represent the state of the Wearable Unit when it has successfully sent over a few character bytes over to the Benchtop unit via the UDP/IP protocol. The Synchronous Reactive director was used as the model of computation because it allows for a signal to have no value at every clock tick. It is suitable for systems which are synchronized such as our Wearable-Benchtop combination. The flat FSM model for the Benchtop unit is embedded inside the "Benchtop Unit" composite actor. There are 5 states in total carrying the same name scheme as shown in Table 1. Each of the states can transition to any one of the four other states depending upon which guard is transitioned in response to receiving a valid (or lack thereof) character byte from the 'Data Stream" actor in every clock tick. The Benchtop unit was found to be fully deterministic which is discussed in more details in Section V.

## IV. RESULTS AND DISCUSSION

Table 3 presents the dataset obtained from the three video clip experiments.

Table 3. Experiment Data

| Self-Report (Clip 1) | Prediction by Wearable (Clip 1) | Self-Report (Clip 2) | Prediction by Wearable (Clip 2) | Self-Report (Clip 3) | Prediction by Wearable (Clip 3) |
|---|---|---|---|---|---|
| NORMAL | NORMAL | HIGH | HIGH | NORMAL | NORMAL |
| NORMAL | NORMAL | NORMAL | MILD | NORMAL | NORMAL |
| NORMAL | NORMAL | NORMAL | MILD | NORMAL | MILD |
| MILD | MILD | HIGH | MILD | MILD | MILD |
| NORMAL | NORMAL | MILD | MILD | NORMAL | MILD |
| NORMAL | NORMAL | HIGH | MILD | NORMAL | NORMAL |
| NORMAL | MILD | NORMAL | NORMAL | NORMAL | MILD |
| MILD | NORMAL | NORMAL | MILD | MILD | NORMAL |
| NORMAL | HIGH | HIGH | MILD | NORMAL | HIGH |
| NORMAL | NORMAL | MILD | MILD | NORMAL | MILD |
| NORMAL | MILD | HIGH | NORMAL | NORMAL | MILD |
| MILD | NORMAL | HIGH | NORMAL | MILD | NORMAL |
| NORMAL | MILD | NORMAL | MILD | NORMAL | MILD |
| NORMAL | MILD | HIGH | NORMAL | NORMAL | MILD |
| NORMAL | NORMAL | HIGH | MILD | NORMAL | MILD |
| NORMAL | NORMAL | HIGH | HIGH | NORMAL | NORMAL |

The entries marked in red indicates the mismatch between the self-reported and predicted class labels. For Video Clip 1, the accuracy is 66.67 % and for Video Clip 2, the accuracy is 13% and for Video Clip 3, the accuracy is 43.75%. The reason for such a wide variation of performance is that emotional response is very subjective and affected by a myriad array of factors. Thus, the self-reported data may have been contaminated by this bias. Furthermore, though each video clips were of same length, the content and their presentation were not "similar" and thus, elicited different emotional cues in the test subject.

Figure 5. demonstrates a simulation which checks the output of the Benchtop FSM model from the inputs of the "Data Stream" actor (simulates Wearable unit).

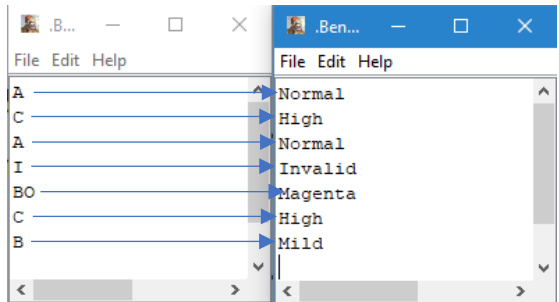

*Fig 5. Simulation of the system*

Note that the "BO" string is not an actual output from the Wearable, but was given here to show that, if there is a 10 second communication gap, the Benchtop unit can transition to the BrownOut state. We further add that, 'BrownOut' event signifies loss of communication between the two unit but if the communication is restored, the system functions as normal as depicted by the last two entries in the "Data Stream" actor. Thus, the Benchtop unit was proved to be fully deterministic.

## V. CONCLUSION AND FUTURE WORK

In this paper we successfully presented a prototype concept of a smart home environment capable of reacting to a user's physiological metrics in real-time without the aid of powerful computers. There were several limitations in our project. Firstly, the PBA algorithm is not optimized to take full advantage of the processing power of 32bit microcontrollers as it was originally written to be used in an 8bit MCUs. Secondly, the GSR data acquisition module lacked proper documentation for which we could not verify if there was any hardware high-pass filter built into the module. Thirdly, we only had one test subject to whose biological signal, the wearable unit was manually calibrated. For future work we suggest a larger test group consisting of people of different ethnicity, age, sex and educational background. Additionally, a model for the Wearable system can be incorporated into the FSM model which would allow for more complex simulations. Finally, a better heartbeat detection algorithm can be employed but care must be taken to ensure that the Heat Rate and GSR readings are synchronized.